\documentclass[twoside,twocolumn,9pt]{article}
\usepackage{extsizes}
\usepackage[super,sort&compress,comma]{natbib} 
\usepackage[version=3]{mhchem}
\usepackage[left=1.5cm, right=1.5cm, top=1.785cm, bottom=2.0cm]{geometry}
\usepackage{balance}
\usepackage{mathptmx}
\usepackage{sectsty}
\usepackage{graphicx} 
\usepackage{lastpage}
\usepackage[format=plain,justification=justified,singlelinecheck=false,font={stretch=1.125,small,sf},labelfont=bf,labelsep=space]{caption}
\usepackage{float}
\usepackage{fancyhdr}
\usepackage{fnpos}
\usepackage{bm}

\usepackage[english]{babel}
\addto{\captionsenglish}{%
  
}
\usepackage{array}
\usepackage{droidsans}
\usepackage{charter}
\usepackage[T1]{fontenc}
\usepackage[usenames,dvipsnames]{xcolor}
\usepackage{setspace}
\usepackage[compact]{titlesec}
\usepackage{hyperref}
\usepackage{tikz}
\usepackage{epstopdf}%This line makes .eps figures into .pdf - please comment out if not required.

\definecolor{cream}{RGB}{222,217,201}

\newcommand{\SH}[1]{\textcolor{black}{#1}}

\begin{document}

\pagestyle{fancy}
\thispagestyle{plain}
\fancypagestyle{plain}{
%%%HEADER%%%
\renewcommand{\headrulewidth}{0pt}
}
%%%END OF HEADER%%%

%%%PAGE SETUP - Please do not change any commands within this section%%%
\makeFNbottom
\makeatletter
\renewcommand\LARGE{\@setfontsize\LARGE{15pt}{17}}
\renewcommand\Large{\@setfontsize\Large{12pt}{14}}
\renewcommand\large{\@setfontsize\large{10pt}{12}}
\renewcommand\footnotesize{\@setfontsize\footnotesize{7pt}{10}}
\makeatother

\renewcommand{\thefootnote}{\fnsymbol{footnote}}
\renewcommand\footnoterule{\vspace*{1pt}% 
\color{cream}\hrule width 3.5in height 0.4pt \color{black}\vspace*{5pt}} 
\setcounter{secnumdepth}{5}

\makeatletter 
\renewcommand\@biblabel[1]{#1}            
\renewcommand\@makefntext[1]% 
{\noindent\makebox[0pt][r]{\@thefnmark\,}#1}
\makeatother 
\renewcommand{\figurename}{\small{Fig.}~}
\sectionfont{\sffamily\Large}
\subsectionfont{\normalsize}
\subsubsectionfont{\bf}
\setstretch{1.125} %In particular, please do not alter this line.
\setlength{\skip\footins}{0.8cm}
\setlength{\footnotesep}{0.25cm}
\setlength{\jot}{10pt}
\titlespacing*{\section}{0pt}{4pt}{4pt}
\titlespacing*{\subsection}{0pt}{15pt}{1pt}
%%%END OF PAGE SETUP%%%

%%%FOOTER%%%
\fancyfoot{}
\fancyfoot[LO,RE]{\vspace{-7.1pt}\includegraphics[height=9pt]{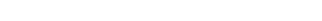}}
\fancyfoot[CO]{\vspace{-7.1pt}\hspace{13.2cm}\includegraphics{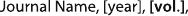}}
\fancyfoot[CE]{\vspace{-7.2pt}\hspace{-14.2cm}\includegraphics{head_foot/RF}}
\fancyfoot[RO]{\footnotesize{\sffamily{1--\pageref{LastPage} ~\textbar  \hspace{2pt}\thepage}}}
\fancyfoot[LE]{\footnotesize{\sffamily{\thepage~\textbar\hspace{3.45cm} 1--\pageref{LastPage}}}}
\fancyhead{}
\renewcommand{\headrulewidth}{0pt} 
\renewcommand{\footrulewidth}{0pt}
\setlength{\arrayrulewidth}{1pt}
\setlength{\columnsep}{6.5mm}
\setlength\bibsep{1pt}
%%%END OF FOOTER%%%

%%%FIGURE SETUP - please do not change any commands within this section%%%
\makeatletter 
\newlength{\figrulesep} 
\setlength{\figrulesep}{0.5\textfloatsep} 

\newcommand{\topfigrule}{\vspace*{-1pt}% 
\noindent{\color{cream}\rule[-\figrulesep]{\columnwidth}{1.5pt}} }

\newcommand{\botfigrule}{\vspace*{-2pt}% 
\noindent{\color{cream}\rule[\figrulesep]{\columnwidth}{1.5pt}} }

\newcommand{\dblfigrule}{\vspace*{-1pt}% 
\noindent{\color{cream}\rule[-\figrulesep]{\textwidth}{1.5pt}} }

\makeatother
%%%END OF FIGURE SETUP%%%

%%%TITLE, AUTHORS AND ABSTRACT%%%
\twocolumn[
  \begin{@twocolumnfalse}
{\includegraphics[height=30pt]{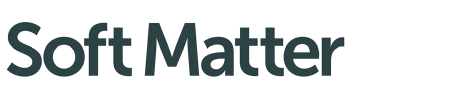}\hfill\raisebox{0pt}[0pt][0pt]{\includegraphics[height=55pt]{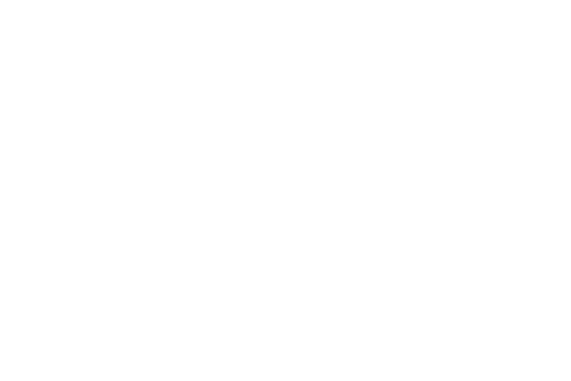}}\\[1ex]
\includegraphics[width=18.5cm]{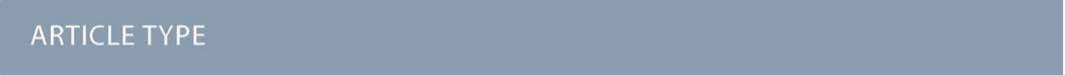}}\par
\vspace{1em}
\sffamily
\begin{tabular}{m{4.5cm} p{13.5cm} }

\includegraphics{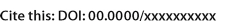} & \noindent\LARGE{\textbf{From a distance: Shuttleworth revisited$^\dag$}} \\%Article title goes here instead of the text "This is the title"
\vspace{0.3cm} & \vspace{0.3cm} \\

 & \noindent\large{Stefanie Heyden$^{\ast}$\textit{$^{a}$} and Nicolas Bain$^{\ast}$\textit{$^{b}$}} \\%Author names go here instead of "Full name", etc.

\includegraphics{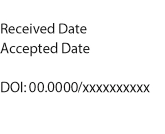} & \noindent\normalsize{
The Shuttleworth equation: a linear stress-strain relation ubiquitously used in modeling the behavior of soft surfaces.
Its validity in the realm of materials subject to large deformation is a topic of current debate.
Here, we allow for large deformation by deriving the constitutive behavior of the surface from the general framework of finite kinematics.
We distinguish cases of finite and infinitesimal surface relaxation preceding an infinitesimal applied deformation. 
The Shuttleworth equation identifies as the Cauchy stress measure in the fully linearized setting.
We show that both in finite and linearized cases, measured elastic constants depend on the utilized stress measure.
In addition, we discuss the physical implications of our results and analyze the impact of surface relaxation on the estimation of surface elastic moduli in the light of two different test cases.
} \\%The abstrast goes here instead of the text "The abstract should be..."

\end{tabular}

 \end{@twocolumnfalse} \vspace{0.6cm}

  ]
%%%END OF TITLE, AUTHORS AND ABSTRACT%%%

%%%FONT SETUP - please do not change any commands within this section
\renewcommand*\rmdefault{bch}\normalfont\upshape
\rmfamily
\section*{}
\vspace{-1cm}

%%%FOOTNOTES%%%

\footnotetext{\textit{$^{a}$~ETH Z\"{u}rich, Institute for Building Materials, 8093 Z\"{u}rich, Switzerland. E-mail: stefanie.heyden@mat.ethz.ch}}
\footnotetext{\textit{$^{b}$~Universite Claude Bernard Lyon 1, CNRS, Institut Lumière Matière, UMR5306, F-69100, Villeurbanne, France. E-mail: nicolas.bain@cnrs.fr }}

%Please use \dag to cite the ESI in the main text of the article.
%If you article does not have ESI please remove the the \dag symbol from the title and the footnotetext below.
\footnotetext{\dag~Electronic Supplementary Information (ESI) available: [details of any supplementary information available should be included here]. See DOI: 10.1039/cXsm00000x/}
%additional addresses can be cited as above using the lower-case letters, c, d, e... If all authors are from the same address, no letter is required

\footnotetext{$\ast$~Both authors contributed equally.}

%%%END OF FOOTNOTES%%%

%%%MAIN TEXT%%%%
\section{Introduction}

\begin{figure}[h]
  \centering
  \includegraphics[scale=1]{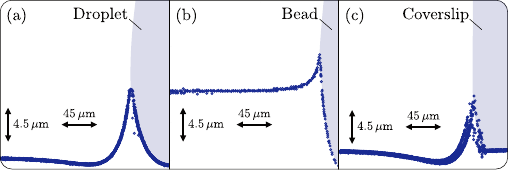}
  \caption{Experimental surface profiles of initially flat soft solids in contact with (a) a drop of glycerol, (b) a silica bead, and (c) a glass coverslip.
    Data were obtained from confocal imaging of fluorescent beads deposited on a silicone gel of shear modulus $\mu\sim 2\,\mbox{kPa}$ (same protocol as \cite{Jagota2012Surface,style2013universal,xu2017direct,xu2018surface}).
    Each blue circle represents a detected fluorescent bead, and the light blue areas represent the contacting objects.}
  \label{fig_examples}
\end{figure}

In most solids, small forces lead to small deformations. 
A soft solid, on the opposite, undergoes large deformations under minute forces: 
a phenomenon as minor as depositing a millimetric droplet leads to large surface deformations~\cite{Style:2017,andreotti2020statics,style2013universal,Hui:2014} (Fig.~\ref{fig_examples}).
Such materials, which can take the form of gels, pastes, or elastomers, are ubiquitous in our lives:
They amount to most of our body tissues, and serve as lubricants, glues, and water-repellent coatings. 
In the past years, it has been shown that surface stresses are essential in wetting, adhesion, and fracture, and can be exploited in composites ~\cite{Style:2017,andreotti2020statics,style2013universal,Hui:2014,Smith2021Droplets,schulman2018surface,hui2020surface,style2013surface,jensen2015wetting,grzelka2017capillary,style2015stiffening}. 
Yet, the physical origins of surface stresses are poorly understood.
This is especially true in gels, in which the cohabitation of a crosslinked polymer network and a liquid solvent complexifies the link between molecular structure and mechanical properties.

The dominant approach to tackle this fundamental question consists in investigating surface elastic properties \cite{Bain:2021,heyden2021contact,Heyden:2021,Carbonaro2020spinningbeads,xu2018surface}.
For instance, surface topography measurements of a stretched patterned silicone gel revealed an elastic surface, where surface stresses increase with surface deformations \cite{Bain:2021}.
This result hints towards a role of the crosslinked polymeric network in the surface constitutive behavior of silicone gels.
Conversely, deformation measures of a spinning hydrogel bead evidenced constant surface stresses, independent of surface deformations, akin to the solvent surface tension \cite{Carbonaro2020spinningbeads}.

In its simplest form, the most common description for the surface mechanics of soft solids relates surface stresses $\bm\sigma_s$ (in N/m) to surface strains $\bm \epsilon_s$ and free energy $W_s$ (in J/m$^2$ or N/m),
\begin{equation}
    \bm \sigma_s = W_s \bm I + \frac{\partial W_s}{\partial \bm \epsilon_s},
    \label{eq:Shuttleworth}
\end{equation}
and is called the Shuttleworth equation \cite{shuttleworth1950surface}.
This description is restricted to the linear regime, where small deformations prevail.
It is, however, largely applied to estimate surface elastic constants (also in N/m) of systems undergoing large deformations, where its validity and resulting physical interpretations have been rightfully questioned \cite{Gutman:2022,Bain:2022,Pandey2020Singular,Masurel2019Disclination}.

\SH{The major drawback of the Shuttleworth equation is to ignore key features that can only be captured by accounting for finite deformations.
Before applied deformations, for instance, soft solids are usually detached from a container, and their surfaces undergo an initial relaxation, inducing residual bulk stresses.
Because soft solids are easy to deform, this surface relaxation can be large \cite{Jagota2012Surface,Mora2013Solid}.}
\SH{Although extensive works in continuum mechanics laid out the framework of surface elasticity (see, e.g., \cite{Gurtin:1975, Gurtin:1998,steigmann1999elastic,Huang:2007,Sharma:2019,Sharma:2020}), employing it in finite element simulations \cite{wu2018effect,liu2020modeling}, sometimes accounting for the surface bending stiffness \cite{steigmann1999elastic,liu2020modeling}, the effect of initial surface relaxations on the estimation of surface elastic constants has so far not been assessed.}
 
\SH{Here, we briefly expose the finite kinematics theory without initial surface relaxation.
We leverage the framework of finite kinematics to rigorously derive surface stress-strain relations from a strain energy density while accounting for prior surface relaxations.
We distinguish infinitesimal and finite surface relaxations and show that, in either case, different stress measures do not coincide, unlike the usual assumption in linear mechanics.
While the classical description Eq.\eqref{eq:Shuttleworth} is valid for the Cauchy stress measure when the surface relaxations are infinitesimal, significant deviation terms appear when they are finite.
We express these deviations in the general case and estimate their magnitude in two test cases.
This framework should incite experimentalists to choose the suitable stress-strain relation and carefully interpret measured surface elastic parameters.}

\section{\SH{Without initial surface relaxations}} 

\subsection*{\SH{A. Finite kinematics}}

\begin{figure}[b]
\begin{centering}
\includegraphics[width = \columnwidth]{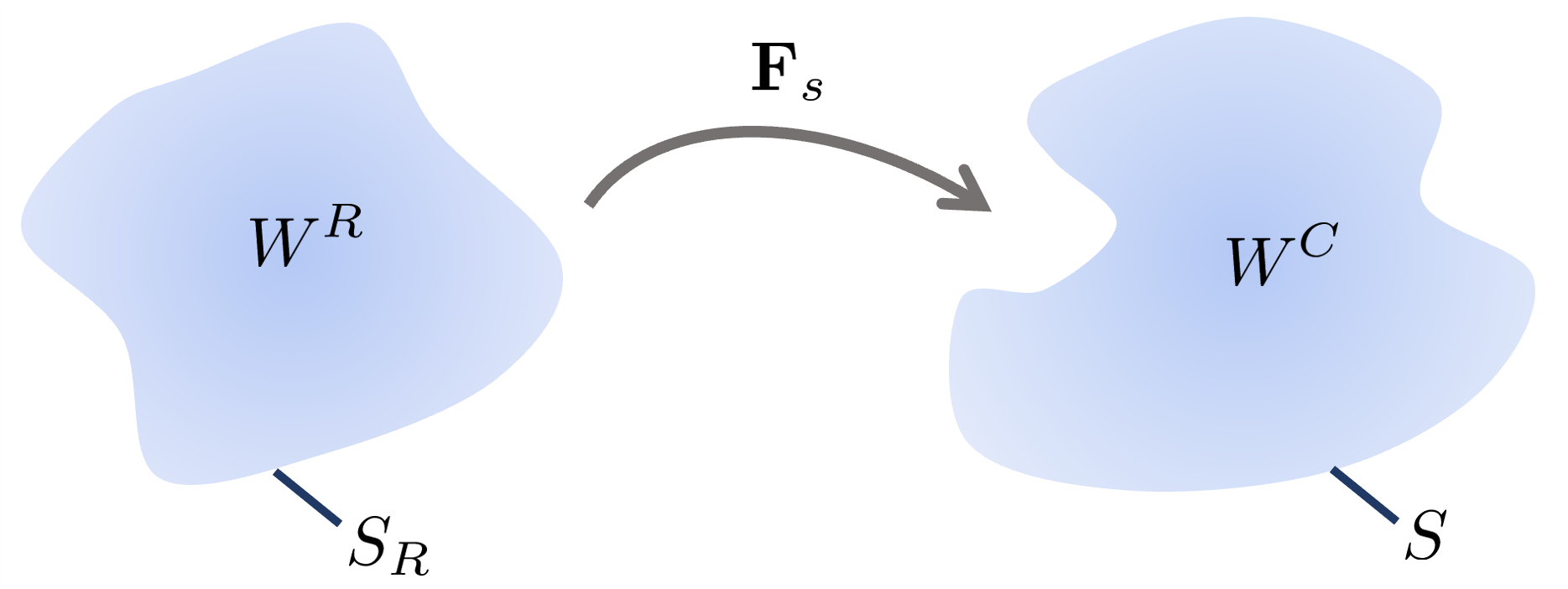}
\end{centering}
\caption{Sketch of reference (left) and deformed configuration (right).}
\label{fig_sketch}
\end{figure}

We start with a brief outline of finite kinematics without initial surface relaxations.
We can consider a solid in two states.
A reference state, before deformation, and a deformed state (Fig.~\ref{fig_sketch}).
In the reference state, the surface exhibits mechanical stresses, which, in the simplest case, have the form of a uniform and isotropic surface tension \cite{Style:2017,andreotti2020statics}.
We furthermore assume that the surface stresses can be thermodynamically defined by a surface strain energy density $W^{\rm \text{\tiny{R}}}$ in the reference configuration.

Following the Piola transform specialized to two dimensions, we define a free energy density expressed in the current configuration $W^{\rm \text{\tiny{C}}}$ as
\begin{equation}
    \int_{S^{\rm \text{\tiny{C}}}}W^{\rm \text{\tiny{C}}}\,dS^{\rm \text{\tiny{C}}} =  \int_{S^{\rm \text{\tiny{R}}}}W^{\rm \text{\tiny{C}}}J_s\,dS_{\rm \text{\tiny{R}}} = \int_{S^{\rm \text{\tiny{R}}}}\,W^{\rm \text{\tiny{R}}}\,dS^{\rm \text{\tiny{R}}},
    \label{eq:ref_current}
\end{equation}
with $S^{\rm \text{\tiny{R}}}$ and $S^{\rm \text{\tiny{C}}}$ the areas in the reference- and current configurations, respectively. 
Furthermore, $J_s=\text{det}(\bm F_s)=\text{det}(\bm I + \bm \nabla_s \bm u)$ measures the local area change, where $\bm F_s$ is the deformation gradient, and $\bm u$ is the displacement field (see Supplementary Section 1).
Based on material frame indifference, any strain energy density can be written as a function of the Green-Lagrange strain tensor $\bm E_s=\frac{1}{2}(\bm F_s^{\rm T}\bm F_s-\bm I)$.
Here, and from now on, all vector quantities, such as displacement fields, and all tensor quantities, such as stress- and strain fields, are projected onto the surface (see Supplementary Section 1).

\subsection*{\SH{B. Different stress measures}}

Depending on the experimental conditions, different stress measures may be used for comparison to theory. 
The first Piola-Kirchhoff stress measure $\bm P_s$, also known as the engineering stress, represents forces in the current configuration per unit referential area.
Therefore, it is the quantity of interest in traction-controlled experiments \cite{Sharma:2020}.
In contrast, the second Piola-Kirchhoff stress measure $\bm S_s$ constitutes forces mapped back to the reference configuration per unit referential area.
Finally, the Cauchy stress measure $\bm\sigma_s$, also denoted as true stress, refers to forces in the current configuration per unit current area.
It is the stress measure used in an Eulerian setting or phenomena where the reference state is ill-defined, as in liquid flows.
By definition, the second Piola-Kirchhoff stress derives from the free energy density in the reference configuration, 
\begin{equation}
    \bm S_s = \frac{\partial W^{\rm \text{\tiny{R}}}}{\partial\bm E_s},
    \label{eq:sec_piola}
\end{equation}
and the two other stress measures account for the change of reference frame by multiplication with the deformation gradient
\begin{equation}
    \bm P_s = \bm F_s \frac{\partial W^{\rm \text{\tiny{R}}}}{\partial\bm E_s} \text{, and }
    \bm \sigma_s = J_s^{-1}\bm F_s\frac{\partial W^{\rm \text{\tiny{R}}}}{\partial\bm E_s}\bm F_s^{\rm T}.
\end{equation}
If one considers the free energy in the current configuration Eq.\eqref{eq:ref_current}, as is implicitly assumed in the Shuttleworth equation, the 
Cauchy stress takes a form similar to Eq.\eqref{eq:Shuttleworth}
\begin{equation}
    \bm \sigma_s = W^{\rm \text{\tiny{C}}}\mathbf I + \bm F_s\frac{\partial W^{\rm \text{\tiny{C}}}}{\partial\bm E_s}\bm F_s^{\rm T},
    \label{eq:Cauchy_Shuttleworth}
\end{equation}
and the other stress measures have a similar expression (see Supplementary Section 2.1).

\subsection*{\SH{C. Surface stress-strain relations}}
It is now instructive to define a constitutive equation for the surface free energy.
Assuming the surface energy is strain-independent $W^{\rm \text{\tiny{C}}} = \gamma$, as in liquids, the Cauchy stresses Eq.~\eqref{eq:Cauchy_Shuttleworth} correspond to an isotropic surface tension $\bm \sigma_s = \gamma\mathbf I$.

For a strain-dependent surface energy, without loss of generality, we use the St. Venant-Kirchhoff model, which is the simplest extension of linear elasticity that captures geometric nonlinearities.
To account for the surface stresses in the reference state, we enrich this model with a constant surface energy in the current configuration,
\begin{equation}
    W^\text{\tiny{R}}(\bm E_s) = \gamma\,J_s + \mu_s\,\text{tr}(\bm E_s\bm E_s) + \frac{1}{2}\lambda_s (\text{tr}(\bm E_s))^2.
    \label{eq:surface_energy}
\end{equation}
Here, $\mu_s$ and $\lambda_s$ are the surface Lam\'{e} parameters.
Physically speaking, $\gamma$ is equivalent to the surface tension of a liquid, and $(\mu_{\rm s}, \lambda_{\rm s})$ are the surface elastic constants attributing an energetic cost to elastic deformation from a reference state.
A straightforward derivation (see Supplementary Section 2.2) leads to the Cauchy stress as a function of the elastic parameters
\begin{equation}
    \bm \sigma_s = \gamma \bm I + 
    \bm F_s \left( 2\mu_s\bm E_s + \lambda_s \text{tr}(\bm E_s)\bm I \right)\bm F_s^T,
\label{eq_cauchy}
\end{equation}
and similarly to the first and second Piola-Kirchhoff stresses
\begin{align}
    \bm P_s &= \gamma J_s\bm F_s \left(\bm F_s^{\rm T}\bm F_s\right)^{-1} + 2\mu_s \bm F_s \bm E_s + \lambda_s \text{tr}(\bm E_s)\bm F_s,\\
    \bm S_s &= \gamma J_s \left(\bm F_s^{\rm T}\bm F_s\right)^{-1} + 
    2\mu_s\bm E_s + \lambda_s \text{tr}(\bm E_s)\bm I.
    \label{eq:second_piola}
\end{align}

\subsection*{\SH{D. Infinitesimal deformations}}

\SH{When surface deformations are finite, the stress measures Eqs.\eqref{eq_cauchy} to \eqref{eq:second_piola} represent different physical quantities and hence differ from one another.
When surface deformations are infinitesimal, however, linearized mechanics do not distinguish between the reference and current states, and all stress measures coincide \cite{anand2020continuum}.
This is the case in the absence of prior surface relaxation: all stress-strains relations give the same constitutive relation upon linearization
\begin{equation}
 \tilde{\bm P_s} = \tilde{\bm S_s} = \tilde{\bm \sigma}_s = \gamma \bm I + 2\mu_s\bm \epsilon_s + \lambda_s \text{tr}(\bm \epsilon_s)\bm I,
\end{equation}
where $\bm\epsilon_s = (\bm\nabla_s\bm u + \bm\nabla_s\bm u^T)/2$ is the linear surface strain, and $(\tilde{\bm P_s},\tilde{\bm S_s},\tilde{\bm \sigma_s})$ are the linearized stress measures.
It is then straightforward to extract surface elastic coefficients when applying small surface deformations \cite{liu2020modeling}.
In the latter, we will show that this property does not hold when we account for surface relaxations before applied deformations.}

\section{\SH{With initial surface relaxations}} 

\subsection*{\SH{A. General setting}}

\begin{figure}[h]
  \centering
  \includegraphics[scale=0.45]{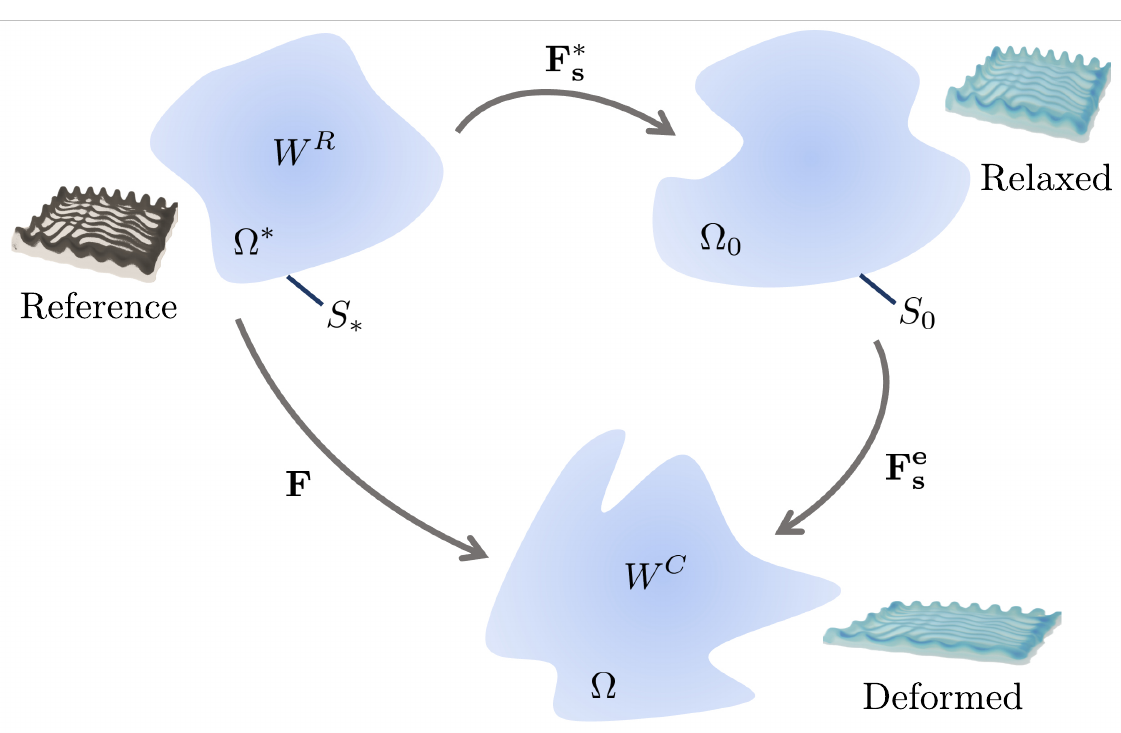}
  \caption{Sketch of fictitious stress-free configuration $\Omega^*$, e.g., a cured sample within a mold, relaxed configuration $\Omega_0$, e.g., a sample after removal from the mold, and deformed configuration $\Omega$.
  The stress-free and the current configurations have a surface strain energy density $W^{\rm \text{\tiny{R}}}$ and $W^{\rm \text{\tiny{C}}}$, respectively.
  }
  \label{fig:sketch}
\end{figure}

Soft solids are often cured within a mold, from which they are detached before being used.
Upon detachment, they undergo a surface relaxation due to an interface change, enhanced wherever the local curvature is nonzero \cite{Jagota2012Surface,Mora2013Solid,hui2020surface}.
We, therefore, distinguish three mechanical states (Fig.~\ref{fig:sketch}). 
The soft solid with the shape of the mold before it relaxes, $\Omega^*$, after it relaxes but before any external load or displacement, $\Omega_0$, and after being deformed by external loads $\Omega$.

In the first state, $\Omega^*$, the bulk of the soft solid is stress-free.
For this reason, we denote this state as the reference configuration (Fig.~\ref{fig_sketch}).
The surface, however, is not stress-free, as is accounted for in our definition of the surface energy Eq.~\eqref{eq:surface_energy}.
While the second state, $\Omega_0$, is the one experimentalists work with, we consider it as an intermediate state because it can contain finite residual bulk stresses due to prior surface relaxation.
We note, however, that these two states coincide when the surface does not undergo prior relaxation, which happens when the soft solid is tested in the state it was prepared.

In the framework of finite kinematics, the deformation gradient $\bm F_s$ maps material points from the stress-free configuration $\Omega^*$ to the deformed configuration $\Omega$.
We also have $\bm F^*_s = \bm I + \boldsymbol{\nabla}_s\bm u^*$ mapping from the stress-free configuration $\Omega^*$ to the relaxed configuration $\Omega_0$, and $\bm F^0_s = \bm I + \boldsymbol{\nabla}_s\bm u^0$ mapping from the relaxed configuration $\Omega_0$ to the deformed configuration $\Omega$.
By composition of mappings, the total deformation follows as $\bm F_s = \bm F^0_s \bm F^*_s$.

In practice, experiments impose deformations $\bm E^0_s$ from the relaxed state $\Omega_0$ to the deformed state $\Omega$.
Here, we focus on the case where imposed deformations are small, for which $\bm E^0_s \sim \bm\epsilon^0_s$.
We then elucidate how the measured stresses vary with imposed surface strain $\bm\epsilon^0_s = [\bm\nabla_s\bm u^0 + (\bm\nabla_s\bm u^0)^T]/2$, both when the relaxation deformation $\bm E^*_s$ is infinitesimal and finite.

\subsection*{\SH{B. Linearized kinematics}}

Let us assume the deformation due to the relaxation of the surface is small $\bm E^*_s \sim \bm\epsilon^*_s$.
Then, the resultant total strain is an additive decomposition $\bm\epsilon_s=\bm\epsilon_s^*+\bm\epsilon_s^0$.

After linearization, stress measures are usually assumed to coincide.
This assumption, however, fails whenever initial stresses are present \cite{Sharma:2019}, as in the case of solid surfaces.
We thus need to carefully distinguish different stress measures in the realm of linearized kinematics.
At first order in strains $\bm\epsilon_s$, the linearized Cauchy stress $\tilde{\bm \sigma}_s$ simplifies to the Hookean form
\begin{equation}
    \tilde{\bm \sigma}_s = \gamma\bm I + 2\mu_s\bm\epsilon_s + \lambda_s\text{tr}(\bm\epsilon_s)\bm I.
\label{eq_cauchy_lin}
\end{equation}
Based on the additive decomposition of strains, Cauchy stresses result in a contribution from the surface relaxation $\tilde{\bm \sigma}_s^*$ and a contribution from the imposed deformations $\tilde{\bm \sigma}_s^0$
\begin{equation}
    \tilde{\bm \sigma}_s = \gamma\bm I + \tilde{\bm \sigma}_s^* + \tilde{\bm \sigma}_s^0,
    \label{eq_cauchy_lin_decomposed}
\end{equation}
where $\tilde{\bm \sigma}_s^\alpha = 2\mu_s\bm\epsilon_s^\alpha + \lambda_s\text{tr}(\bm\epsilon_s^\alpha)\bm I$, with $\alpha\in[*,0]$.
Unsurprisingly, the surface relaxation here comes as an additive stress, as in the classical principle of superposition in linear elasticity.
In this context, the prior surface relaxation does not influence the estimation of the surface elastic moduli $\mu_s$ and $\lambda_s$ from applied strains $\epsilon_s^0$.

Similarly, the second Piola-Kirchhoff stress takes the form
\begin{equation}
    \tilde{\bm S}_s = \gamma\bm I + 2\mu_s^{II}\bm\epsilon_s +  \lambda_s^{II}\text{tr}(\bm\epsilon_s)\bm I,
    \label{eq_sec_piola_lin}
\end{equation}
with effective elastic constants $\mu_s^{II} = \mu_s-\gamma$ and $\lambda_s^{II} = \lambda_s + \gamma$. 
The first Piola-Kirchhoff stress is non-symmetric
\begin{equation}
    \tilde{\bm P}_s = \gamma\bm I + 2\mu_s^{I}\bm\epsilon_s +  \lambda_s^{I}\text{tr}(\bm\epsilon_s)\bm I + \gamma \bm \omega_s,
    \label{eq_first_piola_lin}
\end{equation}
with $\mu_s^{I} = \mu_s-\gamma/2$, $\lambda_s^I = \lambda_s^{II}$, and $\bm\omega_s=(\bm\nabla_s\bm u - \bm\nabla_s\bm u^T)/2$ the infinitesimal rotation tensor (see Supplementary Section 3).
Both the first and second Piola-Kirchhoff stress measures can also be decomposed akin to the Cauchy stress Eq.~\eqref{eq_cauchy_lin_decomposed}.

From Eqs.~\eqref{eq_cauchy_lin} to~\eqref{eq_first_piola_lin}, we note that the three linearized stress measures are only equal in two scenarios.
First, when the solid is unstretched $\bm\epsilon_s = 0$.
All stress measures are then trivially equal to the prestress $\gamma \mathbf I$, which is the surface stress of the solid at rest.
Second, when the prestress is much smaller than the surface elastic constants $\gamma \ll (\mu_{\rm s}, \lambda_{\rm s})$.
In this case, which primarily pertains to hard solids, all stress measures are equal to the Hookean form Eq.\eqref{eq_cauchy_lin}.

\SH{Besides these cases, when the surface moduli is of the order of the surface tension at rest or smaller, we should distinguish the different linearized stress measures. 
Otherwise, the surface shear modulus $\mu_s$ can be misestimated by a value of the order of the surface tension $\gamma$,  Eqs.\,\eqref{eq_sec_piola_lin} and\,\eqref{eq_first_piola_lin}.
This applies to soft solids and complex fluid-fluid interfaces, for which surface tension and surface elasticity can be of the same magnitude \cite{Bain:2021,Heyden:2021,Fuller:2012}.}

\subsection*{\SH{C. Finite kinematics}}

If the surface relaxation induces large deformations, we expand to linear order in $\bm\nabla\bm u^0$ the imposed deformations and keep the relaxation deformations $\bm F^*$ finite.
At first order, we write the dependence of Cauchy stress with imposed deformations
\begin{align}
\bar{\bm \sigma}_s = &\, \gamma\bm I + \bar{\bm \sigma}^*_s + \bm\nabla\bm u^0\bar{\bm \sigma}^*_s + \bar{\bm \sigma}^*_s(\bm\nabla\bm u^0)^T \nonumber\\
&+ 2\mu_s[\bm F_s^*(\bm F_s^*)^T]\bm\epsilon^0_s[\bm F_s^*(\bm F_s^*)^T]\nonumber\\
& + \lambda_s\text{tr}[(\bm F_s^*)^T\bm\epsilon^0_s\bm F_s^*]\bm F_s^*(\bm F_s^*)^T,
\label{eq:finite_relax_cauchy}
\end{align}
where $\bar{\bm \sigma}^*_s = \bm F^*_s \left( 2\mu_s\bm E^*_s + \lambda_s \text{tr}(\bm E^*_s)\bm I \right)(\bm F^*_s)^T$ is the stress contribution due to surface relaxation (see Supplementary Section 3).
Here, not only does the relaxation impose an additional stress term $\bar{\bm \sigma}^*_s$, but it also mixes non-trivially into the terms that include the surface elastic parameters $(\lambda_s, \mu_s)$.
In practice, the exact contribution from the finite surface relaxation depends on the sample geometry and has to be estimated accordingly.
For the sake of completeness, we estimate the strain dependence of the Second Piola-Kirchhoff stress as
\begin{align}
    \bar{\bm S}_s &= \bar{\bm S}_s^* + \gamma J_s^*(\bm F_s^*)^{-1}\left[\text{tr}(\bm \epsilon^0) - 2 \bm\epsilon^0_s\right](\bm F_s^*)^{-T}  \nonumber \\
    &+ 2\mu_s(\bm F_s^*)^T\bm\epsilon^0_s\bm F_s^* + \lambda_s\text{tr}((\bm F_s^*)^T\bm\epsilon^0_s\bm F_s^*)\bm I.
\label{eq:finite_relax_s}
\end{align}
Finally, the First Piola-Kirchhoff stress tensor follows as
\begin{equation}
    \bar{\bm P}_s = \bm F_s^* \bar{\bm S}_s + \bm\nabla_s\bm u^0 \bar{\bm S}_s^*,
\label{eq:finite_relax_p}
\end{equation}
where $\bar{\bm S}_s^* = \gamma J_s^* \left[(\bm F_s^*)^{\rm T}\bm F_s^*\right]^{-1} + 2\mu_s\bm E_s^* + \lambda_s \text{tr}(\bm E_s^*)\bm I$ is the Second Piola-Kirchhoff stress contributions to surface relaxation.
Although the expressions \eqref{eq:finite_relax_cauchy} to \eqref{eq:finite_relax_p} are cumbersome, they do not coincide even in the case of no imposed deformations $\bm\epsilon_s^0 = \bm 0$.

\SH{Overall, whether the surface relaxation induces small or finite deformations, the different stress measures differ from each other.
While we can extract effective surface elastic coefficients in the case of infinitesimal surface relaxation, Eqs.~\eqref{eq_cauchy_lin} to~\eqref{eq_first_piola_lin}, accounting for finite relaxation prevents having a simple constitutive equation between stresses and imposed deformation $\epsilon^0_s$, Eqs.~\eqref{eq:finite_relax_cauchy} to~\eqref{eq:finite_relax_p}.
Still, we approximate from the full expression for the Cauchy stress Eq.\eqref{eq:finite_relax_cauchy} that, at first order in relaxation strain $\epsilon_s^*$, the surface moduli $(\mu_s,\lambda_s)$ will be misestimated by a factor $(1 + 4\epsilon_s^*)$.
Although this suggests that one cannot neglect prior surface relaxations as soon as they reach a few percent, being more precise requires investigating specific test cases.}

\section{\SH{Test cases}} 

\begin{figure}[h!]
\begin{tikzpicture}
    \node (schematic) at (0,8) {\includegraphics[scale=0.25]{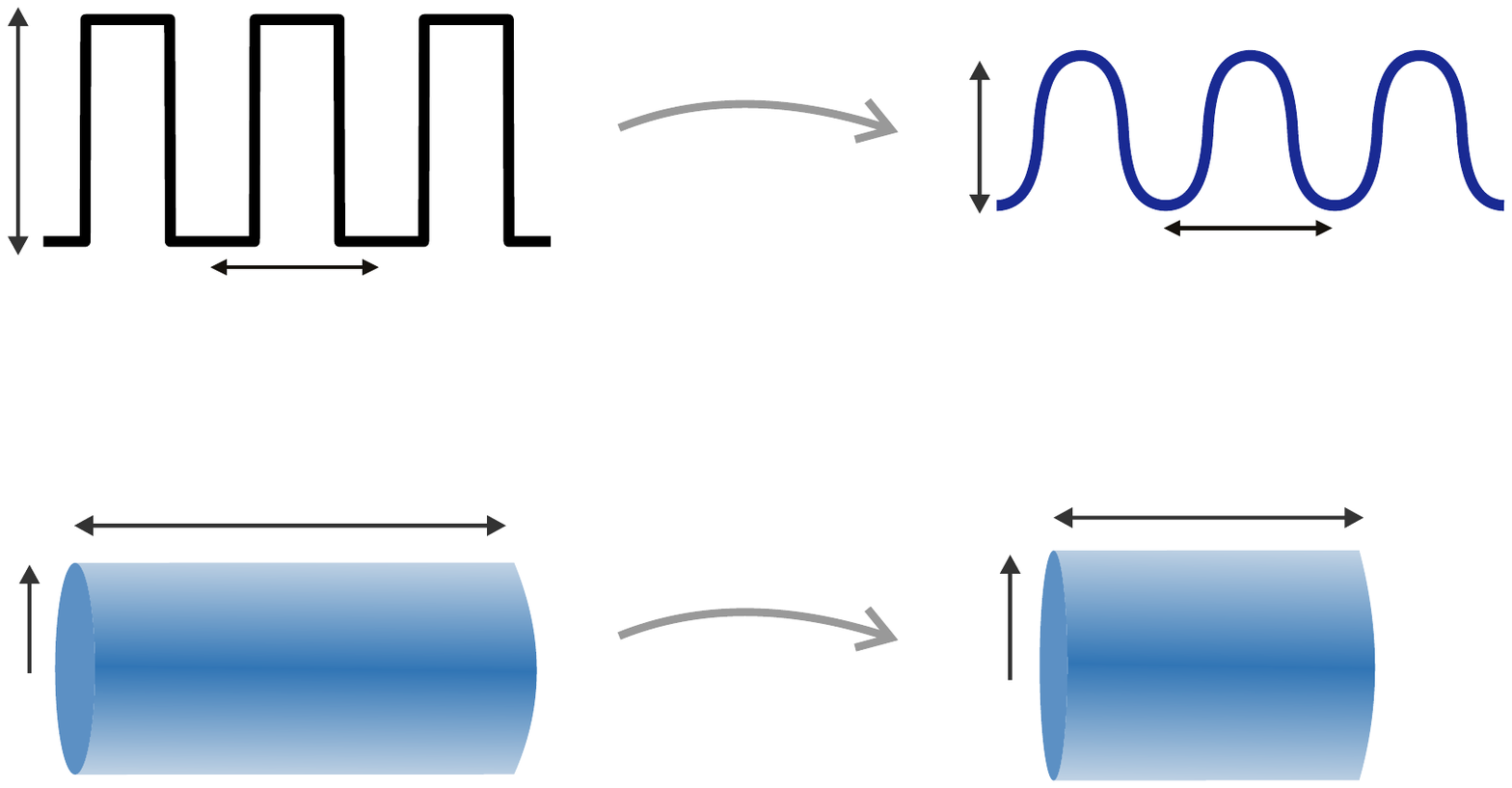}};
    % \node (schematic) at (0,1.5) {\includegraphics[scale=1]{contour_plot.pdf}};
    % \node (schematic) at (0,-4.1) {\includegraphics[scale=1]{corrections_compression.pdf}};
    \node (schematic) at (0,3.6) {\includegraphics[scale=1]{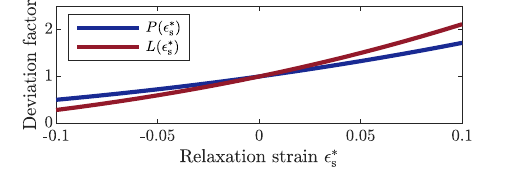}};
    \node (schematic) at (0,-1.8) {\includegraphics[scale=1]{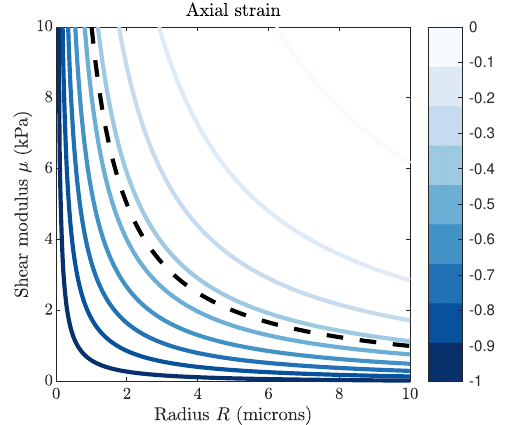}};
    \node[] at (-4.3,9) {(a)};    
    \node[] at (-4.3,6.5) {(b)};
    \node[] at (-4.3,5) {(c)};
    % \node[] at (-4.3,-2.4) {(d)};
    \node[] at (-4.3,1.5) {(d)};
    \node[] at (0,9.4) {$\bm F^*$};
    \node[] at (0,7) {$\bm F^*$};
    \node[] at (-3.6,6.5) {$R$};
    \node[] at (1.1,6.5) {$r$};
    \node[] at (-2.3,7.3) {$L$};
    \node[] at (2.3,7.3) {$l$};
    \node[] at (-3.7,9) {$a_0$};
    \node[] at (-2.1,8.1) {$w$};
    \node[] at (0.9,9.1) {$a_1$};
    \node[] at (2.4,8.3) {$w$};
\end{tikzpicture}
\caption{(a) Surface profile of periodic grooves of wavelength $w$. Left: initial profile, with amplitude $a_0$. Right: after relaxation, with amplitude $a_1$, estimated for a material of shear modulus $\mu = 2\,\rm kPa$, surface tension $\gamma = 20\,\rm mN/m$, wavelength $w = 50\,\mu\rm m$ and initial amplitude $a_0 = 2\,\mu\rm m$.
(b) Cylindrical rod before (left) and after relaxation (right).
(c) Deviation factors $P$ and $L$, quantifying the effect of surface relaxation on estimated surface modulus (Eqs.\eqref{eq_correction_grooves} and \eqref{eq_correction_cylinder}).
(d) Contour plot for the surface relaxation axial strain $\epsilon_{s\parallel}$ as a function of shear modulus and cylinder radius, computed from Eq.\eqref{eq_lambda} with surface tension $\gamma = 20\,\rm mN/m$.
The black dashed line represents stands for $R = L_{\rm ec}$.
}
\label{fig:contours}
\end{figure}

\SH{Let us examine how an incompressible soft solid of bulk shear modulus $\mu$, surface tension $\gamma$, and surface elastic material moduli ($\mu_s$,$\lambda_s$) responds in two canonical examples (Figs.~\ref{fig:contours}a and b).
For both examples, we compute scaling factors to highlight the magnitude by which surface moduli deviate from linear elastic predictions with increasing initial relaxation strain.}

First, we consider a soft solid cured into a mold with periodic rectangular grooves of wavelength $w$ and initial amplitude $a_0$.
For simplicity, we assume that the solid is much thicker than the pattern wavelength, akin to the experimental system in \cite{Bain:2021}.
After demolding, the surface topography relaxes to a nearly sinusoidal wave with final amplitude $a_1 \sim a_0 / (1 + \vert q\vert L_{\rm ec})$, where $q = 2\pi / w$ is the pattern wavevector and $L_{\rm ec} = \gamma / 2\mu$ is the elastocapillary length \cite{Jagota2012Surface,hui2020surface,Bain:2021} (Fig.~\ref{fig:contours}a).
During this process, one period of the surface goes from its initial length $l_0 = w + 2 a_0$ to a final length $l_1 \sim w + 2 a_1$.
We define the surface relaxation strain $\epsilon_s^* = (l_1 - l_0) / l_0$ from the difference in surface length before and after demolding. 

\SH{With no loss of generality, we assume that the surface relaxation strains are tangential to the surface profile $\left(\epsilon_{\rm s\parallel}^*, \epsilon_{\rm s\perp}^*\right) = \left(\epsilon_{\rm s}^*, 0\right)$, without shear, and that we impose an external deformation that manifests as a longitudinal strain $\epsilon_{\rm s\parallel}^0 = \epsilon_{\rm s}^0$ and no transverse applied strain $\epsilon_{\rm s\perp}^0 = 0$.
The longitudinal Cauchy stress,
\begin{equation}
    \bar\sigma_{\rm s\parallel} = \gamma + \bar\sigma_{\rm s\parallel}^* + (2\mu_{\rm s} + \lambda_{\rm s})P(\epsilon_{\rm s}^*)\epsilon_{\rm s}^0,
    \label{eq_correction_grooves}
\end{equation}
is linear in imposed strain $\epsilon_{\rm s}^0$ and polynomial in relaxation strains $P(\epsilon) = (1 + \epsilon)^2\left( 1 + 4\epsilon + 2\epsilon^2\right)$ (see Supplementary Section 4).
When the surface relaxation strain is infinitesimal $\epsilon_{\rm s}^* \ll 1$, we recover the surface modulus $(2\mu_{\rm s} + \lambda_{\rm s})$ that can be calculated from the fully linear Cauchy stress Eq.\eqref{eq_cauchy_lin}.
Otherwise, the estimated surface modulus deviates from the true modulus by a factor $P(\epsilon_{\rm s}^*)$.
This deviation factor increases when the relaxation places the surface in tension, $\epsilon^*_{\rm s} > 0$, and decreases when the surface is in compression $\epsilon^*_{\rm s} < 0$ (Fig.~\ref{fig:contours}c).
In the experimental conditions of \cite{Bain:2021} (recalled in the legend of Fig.~\ref{fig:contours}a), we estimate the relaxation strain to be $\epsilon^*_{\rm s} = -5\%$ and the resulting correction factor $P(-.05) = 0.7$.
Therefore, accounting for the surface relaxation, as done in \cite{Bain:2021} with nonlinear theories, avoided a 30$\%$ underestimation of the surface elastic moduli.}

Second, we assume the soft solid is cured into a slender cylindrical mold of length $L$ and radius $R$, with $L\gg R$ \cite{zafar:2021}.
Once removed from the mold, the length and radius change to $l$ and $r$, respectively (Fig.~\ref{fig:contours}b).
At first order, this relaxation follows the uniform deformation field
\begin{equation}
  \bm F^* = \begin{pmatrix}
    \lambda^* & 0 & 0\\
    0 & \lambda^* & 0 \\
    0 & 0 & \lambda^{*-2}
  \end{pmatrix},
\end{equation}
where $\lambda^*=r/R$ is the radial stretch, and $\lambda^{*-2} = l / L$ the axial stretch.

In this framework, the stretch created by the surface relaxation that minimizes the total elastic energy,
\begin{equation}
    \lambda^* = \left(L_{\rm ec}/R + \sqrt{1 + \left(L_{\rm ec}/R\right)^2}\right)^{1/3},
    \label{eq_lambda}
\end{equation}
depends only on the ratio of elastocapillary length and cylinder radius $L_{\rm ec}/R$ (see Supplementary Section 5).
Here, we denote two types of surface strains: the axial strain $\epsilon_{\rm s\parallel}^* = \lambda^{*-2} - 1$ and the circumferential strain $\epsilon_{\rm s\perp}^* = \lambda^* - 1$.
From Eq.~\eqref{eq_lambda}, surface strains are infinitesimal when the elastocapillary length is much smaller than the cylinder radius, $L_{\rm ec}\ll R$, where bulk elasticity dominates.

Otherwise, the relaxation-induced deformations are finite (Fig.~\ref{fig:contours}d).
If we assume they have no shear component and impose a longitudinal deformation that results in small surface strains $(\epsilon_{\rm s\parallel}^0,\epsilon_{\rm s\perp}^0) = (\epsilon_{\rm s}^0, -\epsilon_{\rm s}^0 / 2)$, the longitudinal Cauchy stress
\begin{equation}
    \bar\sigma_{\rm s\parallel} = \gamma + \bar\sigma_{\rm s\parallel}^* + \left(2\mu_{\rm s} P(\epsilon_{\rm s\parallel}^*) + \frac{\lambda_{\rm s}}{2}L(\epsilon_{\rm s\parallel}^*)\right)\epsilon_{\rm s}^0
    \label{eq_correction_cylinder}
\end{equation}
is linear in imposed strain $\epsilon_{\rm s}^0$ and non-linear in relaxation strains $\epsilon_{\rm s\parallel}^*$ through $P(\epsilon_{\rm s\parallel}^*)$ and the polynomial function
 $L(\epsilon) = (1 + \epsilon)(1 + 8\epsilon + 12\epsilon^2 + 4\epsilon^3)$ (see Supplementary Section 4).
We recover the correct surface modulus $(2\mu_{\rm s} + \lambda_{\rm s} / 2)$ calculated from Eq.\eqref{eq_cauchy_lin} in the case of infinitesimal surface relaxation and increasing deviation factors when surface relaxation is finite (Fig.~\ref{fig:contours}c).
\SH{While the bulk material properties may behave linearly up to large deformations \cite{Bain:2021}, surface elastic constants scale by a factor of 2 in the case of prior surface relaxation on the order of $\pm10\%$ (Fig.~\ref{fig:contours}c).}

\section{Conclusions}

\SH{In this work, we derive the different surface stress-strain relations for a soft solid with and without prior surface relaxations in the regime of finite deformations.
We show that the Shuttleworth equation is only valid for the Cauchy stress without prior relaxations.
It does not apply to other stress measures, and prior relaxations result in deviation factors that lead to misestimating the surface elastic properties.}

\SH{This has direct implications for the determination of surface elastic constants.
First, experimentalists need to assess if the way the solid reached its rest state results in surface relaxations, and if so estimate their amplitude.
Second, they have to evaluate which stress measure is the most relevant depending on how measurements are done.
This calls for a careful notion of which configuration forces and areas are measured.
If forces are measured in the deformed state and areas in the rest state configuration, the first Piola stress measure should be considered, whereas the Cauchy stress should be used when the areas are also measured in the deformed state.
Different constitutive relations need to be employed depending on the stress measure, and the measured elastic moduli will depend on the used stress measure even without prior surface relaxation.}

\SH{As more experimental work is required to determine under which conditions a soft solid has an elastic surface, our results provide a robust framework to interpret measurements of surface elastic constants from different stress measures, whether the surface has or hasn't relaxed during the fabrication process.}

\section*{Author Contributions}

N.B. and S.H. contributed equally to the development of the theory and the writing of the manuscript.
N.B. performed the experimental measurements and data analysis of surface deformations.

\section*{Conflicts of interest}

There are no conflicts to declare.

\section*{Acknowledgements}

S.H. gratefully acknowledges funding via the SNF Ambizione grant PZ00P2186041.
The authors thank Eric R. Dufresne for useful discussions and suggestions.

\balance


%%%REFERENCES%%%
\bibliography{Main_Soft_Matter} %You need to replace "rsc" on this line with the name of your .bib file
\bibliographystyle{rsc} %the RSC's .bst file

\end{document}